\documentclass[aps,prc,onecolumn,amssymb,amsmath,superscriptaddress,floatfix,showpacs,nofootinbib]{revtex4}
\usepackage{graphicx}
\usepackage{dcolumn}
\usepackage{bm}
\begin{document}

\title{{\Large Critical Behavior of J/$\psi$ across the Phase Transition
from QCD Sum Rules}}

\bigskip
\bigskip
\author{\large Kenji Morita}
\email{morita@phya.yonsei.ac.kr}
\affiliation{Institute of Physics and Applied Physics, Yonsei
University, Seoul 120-749, Korea}
\bigskip
\bigskip

\begin{abstract}
\leftskip1.0cm
\rightskip1.0cm
 We study behavior of $J/\psi$ in hot gluonic matter using
 QCD sum rules. Taking into account temperature dependences of the gluon
 condensates extracted from lattice thermodynamics for the pure SU(3)
 system, we find that the mass and width of $J/\psi$ exhibit rapid
 change across the critical temperature.
\end{abstract}

\maketitle

\section{Introduction}

How $J/\psi$ changes its properties in the hot and dense medium is one
of the main interests in high energy heavy ion physics. Pioneering works 
such as a mass shift caused by decreasing string tension
\cite{hashimoto86} and suppression of the yield by Debye screening
\cite{Matsui_PLB178} are based on the fact that heavy quarkonium is a
suitable tool for studying the confinement-deconfinement transition in
QCD. Although experiments reveal things are much more complicated than
expected,
there are plenty of theoretically interesting issues. Especially recent
lattice QCD, based on the maximum entropy method, has shown the spectral
peak can survive even above the critical temperature
\cite{Asakawa_PRL92,Datta_PRD69}, but resolution of the peak is not
sufficient and physics behind the result is still controversial
\cite{umeda_qm08}. 

Motivated by these lattice results, in this work we study the mass and
width of $J/\psi$ in hot gluonic medium using QCD sum rules. QCD sum
rules serve as a reliable non-perturbative theoretical tool for studying
hadrons \cite{Shifman_NPB147}. For charmonium, we can express the
current-current correlation function with the perturbative contribution and the
gluon condensate at the leading order of $\alpha_{s}$ because of much
heavier mass of charm quark than $\Lambda_{\text{QCD}}$. If we set the
energy scale of medium as the same order with $\Lambda_{\text{QCD}}$,
effect of temperature can be entirely described by change of condensates
\cite{Hatsuda93}. Hence, once temperature dependence of relevant
condensates are known, we can calculate properties of the heavy
quarkonium in a quite similar manner to the vacuum case. Here we give a
systematic study of $J/\psi$ using QCD moment sum rules by following the
method described in Ref.~\cite{Reinders_NPB186} with relevant extensions
to the hot gluonic matter case.

\section{Gluon condensate at finite temperature}

\begin{figure}[ht]
 \includegraphics[width=0.45\textwidth]{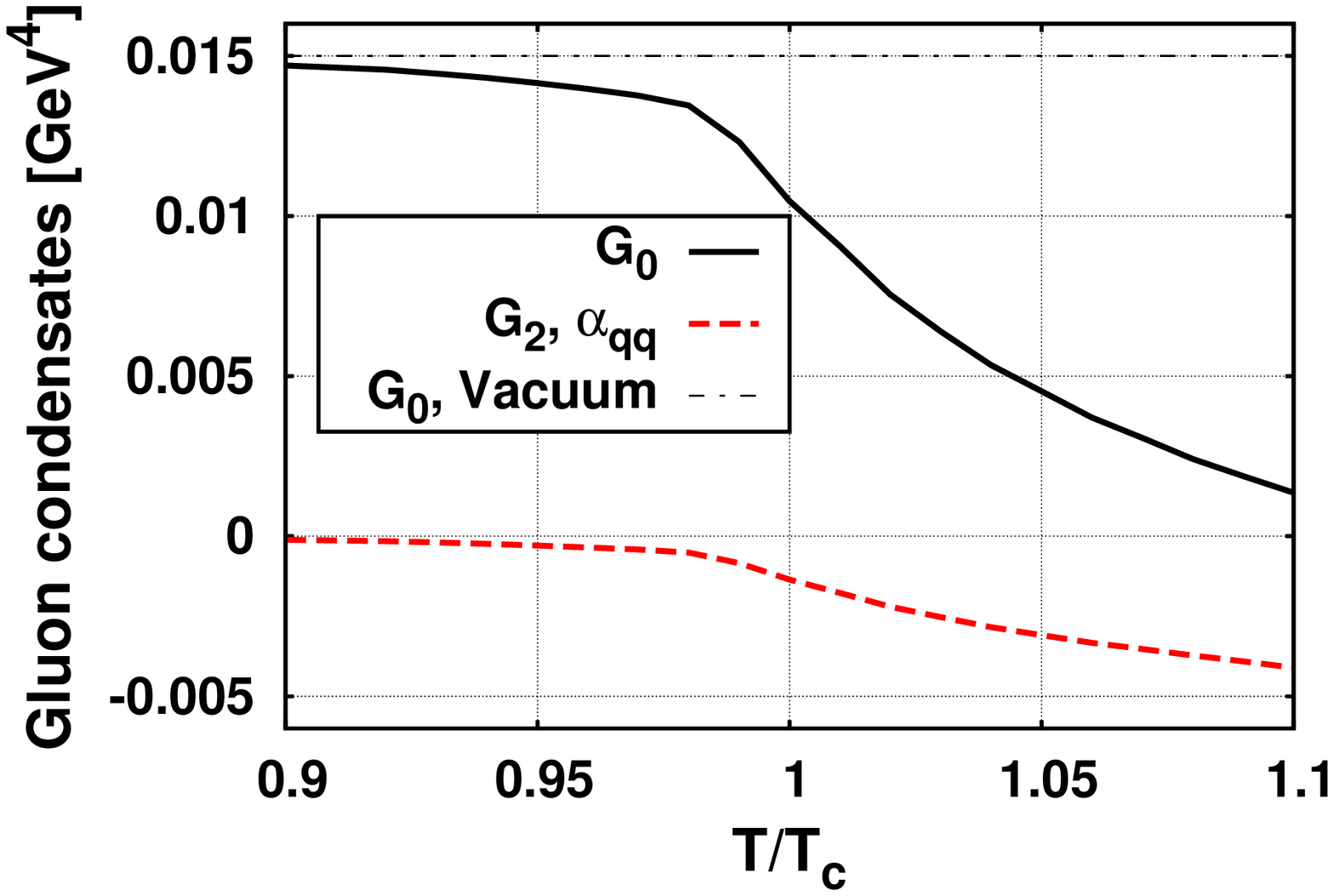}
 \includegraphics[width=0.45\textwidth]{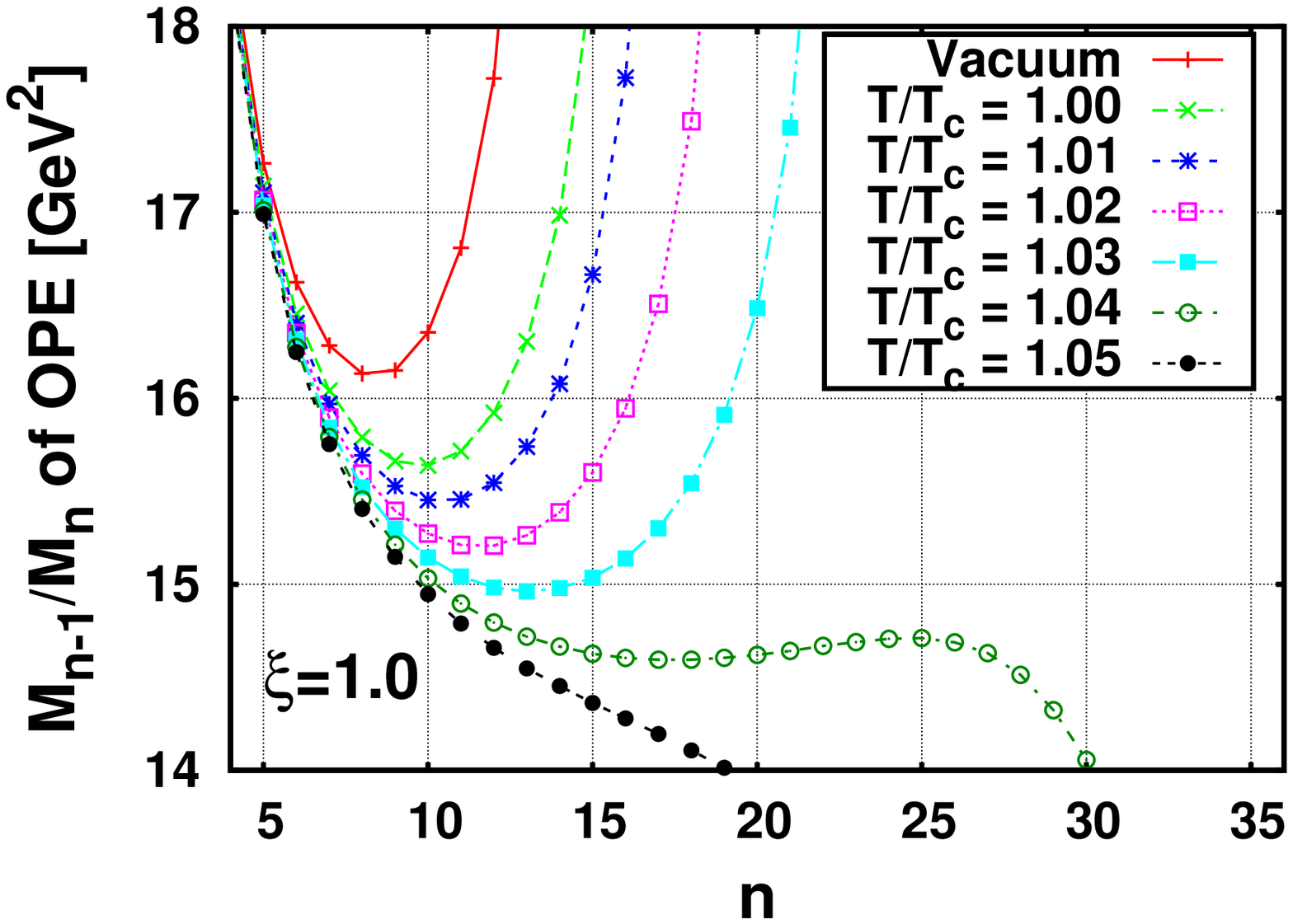}
 \caption{Left: temperature dependence of the gluon condensates. Right:
 ratio of the moment for the OPE side. }
 \label{fig:gc and moment}
\end{figure}

Since in-medium correlation function has no longer Lorentz invariance,
we have to take into account higher twist operators in the operator
product expansion (OPE). In the present case, we consider twist-2 gluon
condensate as well as usual scalar gluon condensate which is related to
the trace anomaly. By decomposing the
twist-2 gluon operator, we define the scalar condensate term $G_0$ and
the traceless and symmetric twist-2 term as
\begin{equation}
 \left\langle \frac{\alpha_s}{\pi}G^a_{\alpha\mu}G_\beta^{a\mu} \right\rangle_T
  = \left(u_\alpha u_\beta -\frac{g_{\alpha\beta}}{4}\right)G_2(T)+G_0(T)
\end{equation}
with $u^\mu$ being the four velocity of the medium. Comparing the
energy-momentum tensor
$T_{\alpha\beta}=(\varepsilon+p)u_{\alpha\beta}-pg_{\alpha\beta}$,
we obtain
\begin{equation}
 G_0(T) =
  G_{0,\text{vac}}-\frac{8}{11}(\varepsilon-3p),
  \quad
  G_2(T) = -\frac{\alpha_s(T)}{\pi}(\varepsilon + p)\label{eq:gc}
\end{equation}
for pure SU(3) system. Here we put the value of the vacuum gluon
condensate to $(0.35\text{GeV})^4$. The three quantities of the
righthand side are extracted from lattice SU(3) calculations
\cite{Boyd_NPB469,Kaczmarek_PRD70}. The resultant condensates are
displayed in the left panel of Fig.~\ref{fig:gc and moment}.

\section{QCD moment sum rules}

We analyzed the spectral property of $J/\psi$ using the moment sum
rules. Definition of the moment of the
correlation function is given by
\begin{equation}
 M_n(Q^2) = \left.\frac{1}{n!}\left(\frac{d}{dq^2}\right)^n \Pi(q^2)\right|_{q^2=-Q^2}\label{eq:moment}
\end{equation}
where $\Pi(q^2)$ is the correlation function for the vector current
\begin{equation}
(q^\mu q^\nu-g^{\mu\nu}q^2) \Pi(q^2) = i\int d^4 x e^{iqx}\langle
 T[\bar{c}(x)\gamma^\mu c(x) \bar{c}(0)\gamma^\nu c(0)] \rangle.
\end{equation}
 The in-medium
correlation function satisfies the dispersion relation
\begin{equation}
 \Pi(q^2) = \int_{0^-}^{\infty}ds
  \frac{\rho(s)}{s-q^2},
\end{equation}
in which $q=(\omega,\boldsymbol{0})$ since we set the
quarkonium at rest with respect to the medium. Taking $q^2 <0$, the
above dispersion relation gives the spectral density $\rho(s)$ in the
righthand side. In the QCD sum rules, we first calculate the
correlation function using OPE which includes the condensates, and then
relate it to the \textit{hadronic} spectral density, which is modeled in
a simple way. Generally the hadronic spectral density contains not only
the desired pole term but also high-energy continuum part. In the
momentum sum rule prescription, we can reduce the contribution from the
continuum so that the pole term dominates the dispersion integral by
optimizing the external parameters $Q^2$ and $n$. Note that since $Q^2=-q^2$
is an external parameter, we can take $q^2$ in the deep spacelike region
such that $\alpha_s(q^2)$ is small enough for the leading order
perturbation.\footnote{$\alpha_s(T)$ in Eq.~\eqref{eq:gc} should be
taken separately because it expresses the effective coupling between
heavy quark and thermalized gluonic medium. }
Hereafter we denote the energy scale in the unit of $4m_c^2$, by
$\xi=Q^2/4m_c^2$.
Following the
optimization criterion used in Ref.~\cite{Reinders_NPB186}, we take the
ratio of the $(n-1)$-th moment to $n$-th one and picking up $n$ at
which the ratio of the OPE side takes its minimum. An example of
this ratio for $\xi=1$ is displayed in the right panel of
Fig.~\ref{fig:gc and moment}. We can see that the ratio
clearly decreases as temperature goes up. However, there is no minimum
at $T=1.05T_{\text{c}}$, that indeed shows breakdown of the theory.
This limitation in the maximally available temperature can be slightly
improved by increasing the scale parameter $\xi$, but current framework
cannot be extended beyond $T=1.06T_{\text{c}}$. See
Ref.~\cite{morita_jpsiprl,morita_jpsifullpaper} for discussion.
From the minima of the ratio, we can extract the mass and width of
$J/\psi$ by applying the relativistic Breit-Wigner form to the
phenomenological hadronic spectral function and equating the ratio,
\text{i.e.}, 
\begin{equation}
 \left.\frac{M_{n-1}}{M_n}\right|_{\text{OPE}} = \left.\frac{M_{n-1}}{M_n}\right|_{\text{phen.}}.
\end{equation}
The moment Eq.~\eqref{eq:moment} enables us to extract to the pole term
at $\xi$ and $n$ determined from the OPE side moment ratio.

\section{Results and discussion}

\begin{figure}[ht]
 \includegraphics[width=0.7\textwidth]{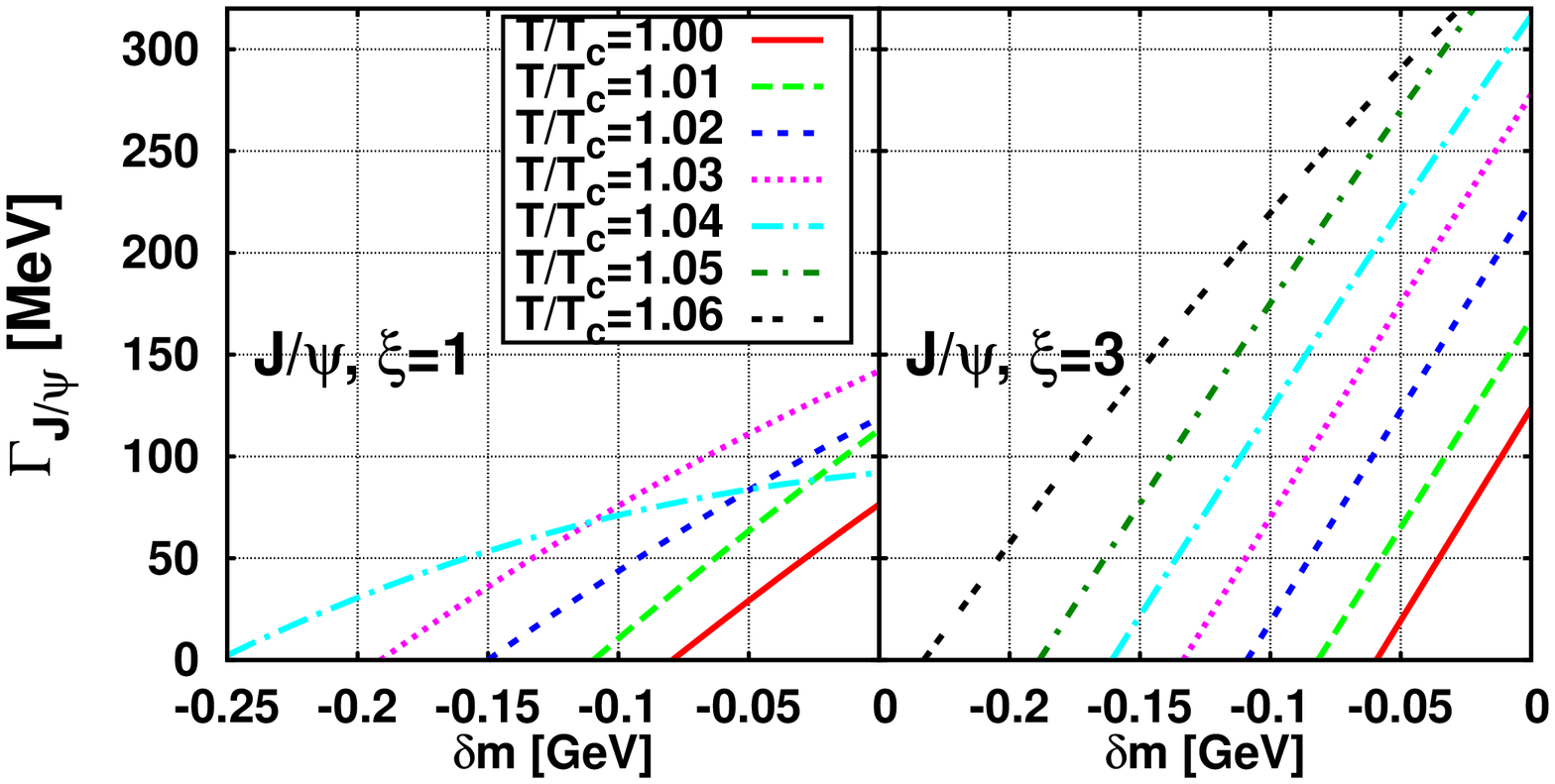}
 \caption{Constraint on the mass shift and width broadening at various
 temperatures. Left panel show the $\xi=1$ case and right panel is for
 $\xi=3$.}
 \label{fig:constraint}
\end{figure}

Figure \ref{fig:constraint} shows the constraint between mass shift and
width obtained from the sum rule. Reflecting the decreasing minimum of
the moment ratio, the change of the mass and width becomes larger as
temperature increases. A notable feature is the
linear relation between the mass and width. Although a deviation from
the linear and parallel behavior among different temperature cases is seen
at $T=1.04T_{\text{c}}$ in $\xi=1$ case, this
can be improved by increasing $\xi$ as shown in the right panel. In fact
the minimum of the ratio obtained at $T=1.04T_{\text{c}}$ is unstable  as
seen in Fig.~\ref{fig:gc and moment}. In such a case, coefficients of OPE
become so large that applicability of the expansion might be suspicious.
Therefore we can conclude that the linear relation holds as far as the
sum rules do. For simultaneous determination of both mass and width, we
need an additional
constraint. See Ref.~\cite{lee_morita_stark} for a recent investigation.
Here we show the possible maximum change of the mass and width, in which the
other quantity is assumed to remain with the vacuum value.

\begin{figure}[ht]
 \includegraphics[width=0.9\textwidth]{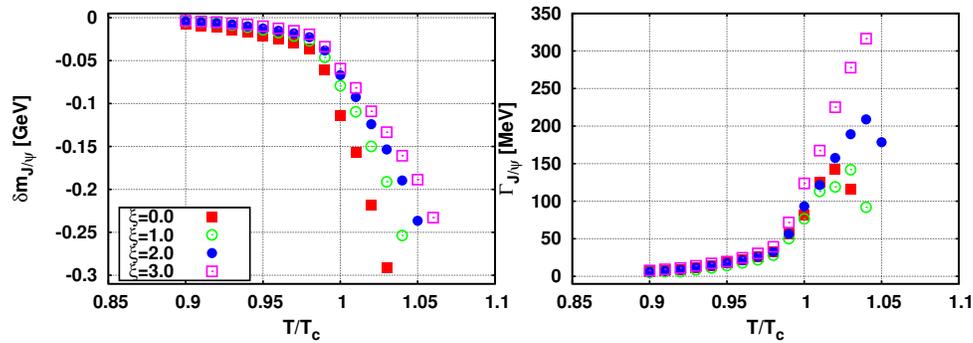}
 \caption{Left: maximum mass shifts for $\Gamma=0$. Right: maximum width
 for $\delta m=0$.}
 \label{fig:shift}
\end{figure}

From Fig.~\ref{fig:shift}, one finds both mass shift and width show
order parameter-like behavior. Since temperature effect is incorporated
into the gluon condensates, which clearly exhibit abrupt change due
to the QCD phase transition, we can conclude the mass and width can be
regarded as ``order parameter'' of the phase transition. Further
discussion and details have been given in
Refs.~\cite{morita_jpsiprl,morita_jpsifullpaper}. 

\section*{Acknowledgement}

The author is much indebted to Prof.~Su Houng Lee for collaboration
throughout this work. This work was supported by BK21 program of Korean
Ministry of Education.



\end{document}